\documentstyle[preprint,aps]{revtex}
\def\mustring{\mu_1...\mu_{p+1}}
\def\alphastring{\alpha_1...\alpha_{p+1}}
\begin{document}
\preprint{\vbox{\hbox{UAHEP-946}\hbox{NI94007}}}
\draft
\title{Black Objects in the Gauge Theory of P-Branes}
\author{B. Harms}
\address{Department of Physics and Astronomy, The University of Alabama,
\\ Tuscaloosa, AL 35487-0324 \\ {\rm and} }
\address{Isaac Newton Institute for Mathematical Sciences,
Cambridge University, \\ Cambridge, CB3 0EH, U.K.}
\author{Y. Leblanc}
\address{Department of Physics and Astronomy, The University of Alabama,
\\ Tuscaloosa, AL 35487-0324}
\maketitle
\begin{abstract}
Within the context of the recently formulated classical gauge theory of
relativistic $p$-branes minimally coupled to general relativity in
$D$-dimensional spacetimes, we obtain solutions of the field equations
which describe black objects.  Explicit solutions are found for two
cases: $D > p+1$ (true p-branes) and $D = p+1$ (p-bags).
\end{abstract}

The correspondence between relativistic strings and Maxwell-type fields
constrained by the condition $\epsilon_{\alpha\beta\mu\nu}
F^{\alpha\beta} F^{\mu\nu} = 0$ was established many years ago
\cite{nam,kas}.  The relations between the string models and the gauge
field description are established by the so-called Pl\" ucker's coordinates
\cite{nam,kas}.

Recently this correspondence has been extended to the case of higher
dimensional objects ($p$-branes) \cite{aur}.  For membranes ($p = 2$)
the corresponding gauge field description is in terms of Kalb--Ramond-type
fields \cite{gre}.  The generalization to arbitrary $p$ was also treated in
Ref.\cite{aur}.  In the general case the curvature of the corresponding gauge
field carries $p+1$ indices, each running over the full spacetime
dimensionality $D$.  We first briefly review the relevant results of
Ref.\cite{aur}, and then show how singular solutions of the field equations
arise in this formalism.

The theory of relativistic closed $p$-branes minimally coupled to general
relativity (GR) is well known to be described by the generalized Nambu-Goto
action,
\begin{eqnarray}
S_{NG} = -\rho_p \int d^{p+1} \sigma \Vert \dot{X} \Vert \; ,
\end{eqnarray}
in which
\begin{eqnarray}
\dot{X}^{\mu_1,...,\mu_D} \equiv \partial_1 X^{\mu_1}\wedge ... \wedge
\partial_{p+1}X^{\mu_{p+1}} \; \ \  ; \mu_i = 0,1,..., D-1 \; ,
\end{eqnarray}
is the tangent vector to the world hypersurface,
\begin{eqnarray}
\Vert \dot{X} \Vert^{2} = {-1 \over{(p+1)!}} \dot{X}^{\mu_1...\mu_{p+1}}
\dot{X}^{\nu_1...\nu_{p+1}} g_{\mu_1\nu_1}... g_{\mu_{p+1}\nu_{p+1}} \; ,
\end{eqnarray}
where
$X^{\mu}(\sigma)$ are the $p$-brane coordinates (embeddings) and
$g_{\mu\nu}$ is the spacetime metric. $\rho_p$ is the hypersurface tension.

In Ref.\cite{aur} the above theory is shown to be equivalently described by
the following (classical) gauge theory spacetime action
\begin{eqnarray}
S &=& - \lambda^2 \int_{\cal M} d^Dx \sqrt{-g}\; \Vert W(x)\Vert +
{1\over{(p+1)!}} \int_{\cal M} d^D x \sqrt{-g}\; W(x)\cdot F(x)\nonumber \\
&& - {1\over{16\pi G_N}}\int_{\cal M} d^D x \sqrt{-g} R(x) \; ,
\end{eqnarray}
where ${\cal M}$ is the $D$-dimensional spacetime manifold, and
$ F_{\mu_1...\mu_{p+1}} = \partial_{[\mu_1}B_{\mu_2...\mu_{p+1}]}$
is the totally antisymmmetric gauge field strength with potential
$B_{\mu_2...\mu_{p+1}}$. $W_{\mu_1...\mu_{p+1}}$ is an auxiliary field
related to the $p$-brane coordinates as follows \cite{aur}
\begin{eqnarray}
W^{\mu_1...\mu_{p+1}} &=& {m\over{\sqrt{-g}}} \int d^{p+1}\sigma
\delta^{(D)}(x-X(\sigma)) \dot{X}^{\mu_1...\mu_{p+1}}(\sigma) \; ,\nonumber \\
&=& m J^{\mu_1...\mu_{p+1}} \; ,
\end{eqnarray}
where $m$ is a constant with dimension $L^{D-p-3}$ and
$J^{\mu_1...\mu_{p+1}}$ is the $p$-brane current.

Varying the action (4) with respect to $W, B$ and $g$, we arrive at the
following field equations,
\begin{eqnarray}
\partial_{\mu_1}[\sqrt{-g}W^{\mustring}] = 0 \; , \\
F_{\mustring} = -\lambda^2{ W_{\mustring}\over{\Vert W\Vert}} \; ,
\end{eqnarray}
and
\begin{eqnarray}
R_{\mu\nu} - {1\over{2}} g_{\mu\nu} R = 8\pi G_N T_{\mu\nu} \; ,
\end{eqnarray}
in which we defined
\begin{eqnarray}
T_{\mu\nu} &\equiv& -2{\delta {\cal L}_{\rm matter}\over{\delta
g^{\mu\nu}}} + g_{\mu\nu} {\cal L}_{\rm matter} \nonumber \\
&=& {\lambda^2\over{p!}}{W_{\mu}^{\mu_2...\mu_{p+1}}
W_{\nu\mu_2...\mu_{p+1}}\over{\Vert W \Vert}} \; .
\end{eqnarray}
Notice that to arrive at Eq.(9), use has been made of the field equation (7).
Using Eq.(5) for the $p$-brane current, the on-shell energy-momentum tensor
$T_{\mu\nu}$ of the $p$-brane can be rewritten as,
\begin{eqnarray}
T_{\mu\nu}(x) &=& {\lambda^2 m \over{\sqrt{-g}}}
{J_{\mu}^{\mu_2...\mu_{p+1}}(x) \; J_{\nu\mu_2...\mu_{p+1}}(x)
\over{\Vert J \Vert}} \nonumber \\
&=& {\lambda^2 m \over{\sqrt{-g}}}\int d^{p+1} \sigma
{\dot{X}_{\mu}^{\mustring} \dot{X}_{\nu\mustring}
\over{\Vert \dot{X} \Vert}} \delta^{(D)}(x - X(\sigma)) \;
\end{eqnarray}
displaying the equivalence of both descriptions \cite{aur}.

A straightforward calculation \cite{aur} shows that the $p$-brane
energy-momentum tensor conservation is equivalent to the $p$-brane equations
of motion,
\begin{eqnarray}
\nabla^{\mu} T_{\mu\nu} &=& 0 \; , \nonumber\\
&\Leftrightarrow& [m,2,...,p+1]\; \partial_m
\Bigl[{\partial_2 X^{\mu_2} ... \partial_{p+1} X^{\mu_{p+1}}
\dot{X}_{\nu\mu_2...\mu_{p+1}} \over{\Vert \dot{X} \Vert}} \Bigr] = 0 \; ,
\end{eqnarray}
where $[1, ..., p+1]$ is the totally antisymmetric tensor in the world
hypersurface of the $p$-brane.

Our goal is to find singular spacetime solutions to the field equations
(6)-(8), and as we show below it is possible to extract such solutions from
the gauge theory formulation of $p$-branes.

We assume the following ansatz for the $D$-dimensional spacetime metric
$(D > p+1)$,
\begin{eqnarray}
ds^2 = -{\rm e} ^{2\Phi(r)} dt^2 + {\rm e}^{-2\Phi(r)} dr^2 +
r^2 d\Omega^2_{p-1} + \sum_{i=p+1}^{D-1} dx^2_i \; ,
\end{eqnarray}
so that,
\begin{eqnarray}
g_{\mu\nu} = \left( \begin{array}{cc} \tilde{g}_{\alpha\beta} \ \  0 \\
0 \ \ \delta_{jk} \end{array}\right) \; ; \ \
\left(\begin{array}{c} \alpha , \beta = 0,1,...,p
\\ j,k = p+1,...,D-1\end{array}\right) \; .
\end{eqnarray}
Clearly this ansatz describes a spacetime composed of the direct sum of a
$p+1$-dimensional Schwarzschild--de Sitter spacetime and a
$(D-p-1)$-dimensional flat Minkowski space.  For a spacetime partitioned in
this way a solution of Eq.(6) can be constructed as follows,
\begin{eqnarray}
W^{\mustring} &=& {-c\over{\sqrt{-\tilde{g}}}}  [\mustring] \; ;
\ \ ( \mu_i = 0,...,p )\; ; \nonumber \\
& =& 0 \; ; \ \ (\mu_i > p) \; ,
\end{eqnarray}
in which the totally antisymmetric tensor is defined as ,
\begin{eqnarray}
[\alpha_1,...,\alpha_{p+1}] \equiv \left\{ \begin{array}{c} +1 \ \
{\rm even \; perm.\; of}\; \alpha_1,...,\alpha_{p+1} \\
-1 \ \ {\rm  odd \; perm.\; of}\; \alpha_1,...,\alpha_{p+1} \\
0 \ \ {\rm otherwise\ \ \ \ \ \ \ \ \ \ \ \ \ \ \ \ \ } \end{array}\right\}
\;  .
\end{eqnarray}
In curved spacetime the Levi-Civita tensor is given as \cite{mis}
\begin{eqnarray}
\epsilon_{\alphastring} = \sqrt{-\tilde{g}}\; [\alphastring] \nonumber \\
\epsilon^{\alphastring} = {-1\over{\sqrt{-\tilde{g}}}} \; [\alphastring] \; .
\end{eqnarray}
Since $\det\vert g_{\mu\nu}\vert = g = \det \vert\tilde{g}_{\alpha\beta}
\vert = \tilde{g}$, we have
\begin{eqnarray}
W^{\mustring} &=& c\epsilon^{\mustring} \; ; \ \ \mu_i = 0,...,p \nonumber \\
&=& 0 \; \ \ \mu_i > p \nonumber \\
F^2 &=& -\lambda^2 (p+1)! \; .
\end{eqnarray}
The Levi--Civita tensor satisfies the relation
$\epsilon^{\alpha\mu_2...\mu_{p+1}}\epsilon_{\beta\mu_2...\mu_{p+1}}
= -p! \delta_{\beta}^{\alpha}, (\alpha, \beta = 0,...,p)$,
so the $p$-brane energy-momentum tensor is
\begin{eqnarray}
T_{\mu\nu} &=& -\lambda^2 c g_{\mu\nu}\; ;\ \ \  (\mu,\nu = 0,...,p)
\nonumber \\
&=& 0 \ \  (\mu,\nu > p) \; .
\end{eqnarray}
The Einstein field equations now read
\begin{eqnarray}
R_{\mu\nu} - {1\over{2}} g_{\mu\nu} R
&=& - \Lambda g_{\mu\nu} \; ; \ \ (\mu,\nu = 0,...,p) \nonumber \\
&=& 0 \; ;  \ \ (\mu, \nu > p) \; ,
\end{eqnarray}
in which the classical $p$-brane induced cosmological constant $\Lambda$ is
given as
\begin{eqnarray}
\Lambda = 8\pi G_N \lambda^2 c \; .
\end{eqnarray}
A singular solution of Eq.(19) is the Schwarzschild -- de Sitter black hole
in $(p+1)$ dimensions,
\begin{eqnarray}
g_{00} = -{\rm e}^{2\Phi(r)} = -\Bigl[1 - {\kappa \over{r^{p-2}}} -
{2\Lambda r^2 \over{p(p-1)}} \Bigr] \; .
\end{eqnarray}
The parameters $\kappa$ and $\Lambda$ determine the position of the two
horizons which are present in such a spacetime.  However, because the
solution is actually embedded in $D$-dimensional spacetime with a flat
$(D-p-1)$-dimensional Minkowski subspace, it actually describes a black
$(D-p-1)$-brane in $D$ dimensions.  Black $p$-brane solutions have previously
been constructed in gauge theories (such as the Kalb--Ramond field) minimally
coupled to general relativity \cite{gar}.  Given the present (classical)
equivalence between the $p$-brane and gauge field formulations, it is
therefore not surprising, although reassuring, to be able find black
$p$-branes in the latter formulation.

In the analysis above we have stipulated that $D > p+1$.  The case $D = p+1$
($p$-bags) must be treated seperately.  The hypersurface dimensionality of
$p$-bags is the same as the dimensionality of the spacetime in which it is
embedded and so the embeddings $X(\sigma)$ are pure general coordinate
transformations.  By definition a $p$-bag is an open object \cite{aur}, so
the gauge theory action Eq.(4) is no longer complete.  There is an additional
term which takes into account the coupling of the gauge potential to the
boundary current.  The term which must be added to Eq.(4) is of the form
\begin{eqnarray}
S_{\rm bd} &=& {f\over{(D-1)!}}\int_{\cal M} d^Dx \sqrt{-g} B_{\mu_2...\mu_D}
J^{\mu_2...\mu_D} \nonumber \\
&& - {1\over{2\pi\alpha_D'}} \int_{\partial U} d^{D-1}\sigma
\sqrt{{-1\over{(D-1)!}} \dot{X}^{\mu_2...\mu_D} \dot{X}_{\mu_2...\mu_D}} \; ,
\end{eqnarray}
in which $J^{\mu_2...\mu_D}$ is the boundary current with coupling strength
$f$ and $\partial U $ stands for the boundary of the $p$-bag $(p = D-1)$.
The matter field equations are now
\begin{eqnarray}
\partial_{\mu_1} [\sqrt{-g} W^{\mustring}] &=& {-f\over{(D-1)!}}
J^{\mu_2...\mu_D} \\
{-\lambda^2 W_{\mustring}\over{\Vert W \Vert}} &=& F_{\mustring} \; ,
\end{eqnarray}
and
\begin{eqnarray}
[1,...,D-1] \; \partial_1 \Bigl[\dot{X}_{\mu\mu_3...\mu_D}
{\partial_2 X^{\mu_3}...\partial_{D-1} X^{\mu_D}\over{\Vert \dot{X} \Vert}}
\Bigr] = {-f\over{(D-1)!}} F_{\mu\mu_2...\mu_D} \dot{X}^{\mu_2...\mu_D} \; .
\end{eqnarray}
The solution of the field equations (Eqs. (23),(24)) are then \cite{aur}
\begin{eqnarray}
W^{\mustring} = \epsilon^{\mustring} [c - f\Theta_U(x)] \; ,
\end{eqnarray}
and
\begin{eqnarray}
F_{\mustring} = \lambda^2 \epsilon_{\mustring} \; ,
\end{eqnarray}
in which $\Theta_U(x)$ is a generalized step function defined as
\begin{eqnarray}
\Theta_U(x) &=& 1, \ \ x \; {\rm inside} \;  \partial U \nonumber \\
            &=& 0, \ \ x\;  {\rm outside} \; \partial U \; .
\end{eqnarray}
Inserting the solutions Eqs.(28)-(29) into the full action yields GR with a
cosmological constant \cite{aur}
\begin{eqnarray}
S_{\rm total} &=& {-1\over{2\pi\alpha_D'}}\int_{\partial U} d^{D-1}\sigma
\sqrt{{-1\over{(D-1)!}}\dot{X}^{\mu_2...\mu_D}
\dot{X}_{\mu_2...\mu_D}}\nonumber \\
&&-{1\over{16\pi G_N}}\int_{\cal M} d^D x \sqrt{-g}
\Bigl[ R(x) - 2\Lambda \Theta_U(x) \Bigr] \; .
\end{eqnarray}
The cosmological constant in the expression above is non-vanishing only in
the interior region of the $p$-bag and is related to the boundary coupling
strength by
\begin{eqnarray}
\Lambda = -8\pi G_N f\lambda^2 \; .
\end{eqnarray}
As for the $p$-brane the $p$-bag system possesses spherically symmetric,
$D$-dimensional black hole solutions.  In the region exterior to the $p$-bag
the solution is the usual Schwarzschild black hole
\begin{eqnarray}
g_{00}^{out} = -\Bigl[ 1 - {\kappa \over{r^{D-3}}}\Bigr] \; ; \ \ r>r_U \; ,
\end{eqnarray}
where $r_U$ is the position of the boundary.  The interior solution has a
cosmological constant, so the solution is Schwarzschild--de Sitter
\begin{eqnarray}
g_{00}^{in} = \Bigl[ 1 - {\kappa'\over{r^{D-3}}} - {2\Lambda r^2
\over{(D-1)(D-2)}}\Bigr] \; ; \ \ r<r_U  \; .
\end{eqnarray}
These solutions are readily derived from, e.g., the $(0,0)$-component of the
Einstein field equations
\begin{eqnarray}
\partial_{r}^2 g_{00} + {(D-2)\over{r}} \partial_r g_{00} -
{ 4\Lambda\over{D-2}} \Theta_U(r) = 0 \; .
\end{eqnarray}
Continuity of the metric at the p-bag boundary requires the relationship
\begin{eqnarray}
g^{in}(r=r_U) = g^{out}(r=r_U) \; ,
\end{eqnarray}
from which we find
\begin{eqnarray}
\kappa = \kappa' + {2\Lambda r_U^{D-1} \over{(D-1)(D-2)}} \; .
\end{eqnarray}
This relation shows that observers inside the bag will measure a different
mass for the black hole than that measured by observers outside the bag.  The
cosmological constant due to the bag boundary generates a mass
``renormalization''.  In this system the metric tensor is continuous across
the boundary, but the curvature tensor is not.  Notice finally the confining
property of the gravitational potential inside the bag.

The generalization of the results presented here to the case of charged and
dilatonic black holes can be obtained by adding the appropriate terms
\cite{gib} to the action in Eq.(4).

\acknowledgments

One of the authors (B.H.) would like to thank the Isaac Newton Institute of
Mathematical Sciences for its hospitality while this work was being carried
out.
This work was supported in part by the U.S. Department of Energy under Grant
No. DE-FG05-84ER40141.


\begin{references}

\bibitem{nam}Y. Nambu, Phys.\ Rev. {\bf D10}, 4262(1974); Phys.\ Lett.
{\bf 92B}, 327(1980); Phys.\ Lett. {\bf 102B}, 149(1981).
\bibitem{kas}H.A. Kastrup, Phys.\ Lett. {\bf 78B}, 433(1978);{\bf 82B},
237(1979); Phys.\ Rep. {\bf 101}, 1(1983); H.A. Kastrup and M.A. Rinke,
Phys.\ Lett. {\bf 105B}, 191(1981); M.A. Rinke, Comm. \ Math. \ Phys.
{\bf 73}, 265(1980),
\bibitem{aur}A. Aurilia and E. Spalluci, Trieste Report No. UTS--DFT-92-5,
1994.
\bibitem{gre}M.B. Green, J.H. Schwarz and E. Witten, {\it Superstring Theory,
Vol. I}, (Cambridge University Press, Cambridge, 1987); and references
therein.
\bibitem{mis}C.W. Misner, K.S. Thorne and J.A. Wheeler, {\it Gravitation},
(W.H. Freeman and Co., San Francisco, 1973).
\bibitem{gar}D. Garfinkle, G.T. Horowitz and A. Strominger, Phys.\ Rev.
{\bf D43}, 3140(1991).
\bibitem{gib}G.W. Gibbons and Kei-ichi Maeda, Nuc.\ Phys. {\bf B298},
741(1988).
\end{references}
\end{document}